\documentclass[12pt]{article}

\usepackage{amsfonts}
\usepackage{amsmath}
\usepackage{epsfig}
\usepackage{latexsym}
\usepackage{graphicx}
\numberwithin{equation}{section}
\allowdisplaybreaks

\def\spa#1{\phantom{\fbox{\rule[-#1cm]{0cm}{0cm}}}}

\def\be{\begin{equation}}
\def\ee{\end{equation}}
\def\bea{\begin{eqnarray}}
\def\eea{\end{eqnarray}}

\def\half{{1\over 2}}
\def\Tr{\mbox{Tr}}
\def\del{\partial}
\def\nn{\nonumber}

\renewcommand{\thefootnote}{\fnsymbol{footnote}}

\begin{document}

\hfuzz=100pt
\title{{\Large \bf{Summing Up All Genus Free Energy\\ of ABJM Matrix Model}}}
\date{}
\author{Hiroyuki Fuji$^a$\footnote{fuji@th.phys.nagoya-u.ac.jp},\;
Shinji Hirano$^b$\footnote{hirano@eken.phys.nagoya-u.ac.jp}, and 
Sanefumi Moriyama$^c$\footnote{moriyama@math.nagoya-u.ac.jp}
  \spa{0.5} \\
$^{a,b}${{\it Department of Physics}}
\\ {{\it Nagoya University}}
\\ {{\it Nagoya 464-8602, Japan}}
  \spa{0.5} \\
$^c${{\it Kobayashi Maskawa Institute}}
\\ {{\it Nagoya University}}
\\ {{\it Nagoya 464-8602, Japan}}
}
\date{}

\maketitle
\centerline{}

\begin{abstract}
The localization technique allows us to compute the free energy of the
$U(N)_k\times U(N)_{-k}$ Chern-Simons-matter theory dual to type IIA
strings on $AdS_4\times CP^3$ from weak to strong 't Hooft coupling
$\lambda=N/k$ at finite $N$, as demonstrated by Drukker, Mari\~no, and
Putrov.
In this note we study further the free energy at large 't Hooft
coupling with the aim of testing AdS/CFT at the quantum gravity level
and, in particular, sum up all the $1/N$ corrections, apart from the
worldsheet instanton contributions.
The all genus partition function takes a remarkably simple form -- the
Airy function,
$\mbox{Ai}\bigl((\pi k^2/\sqrt{2})^{2/3}
\lambda_{\rm ren}\bigr)$,
with the renormalized 't Hooft coupling $\lambda_{\rm ren}$. 
\end{abstract}

\renewcommand{\thefootnote}{\arabic{footnote}}
\setcounter{footnote}{0}

\newpage

\section{Introduction and main results}
There has been considerable progress in testing the AdS/CFT conjecture
\cite{Maldacena:1997re} in the large $N$ limit.
Through its connection to the spin chain systems, the large $N$
integrability allowed us to successfully check the duality from weak
to strong 't Hooft coupling with remarkably high accuracy
\cite{Beisert:2010jr}.
In the meantime, there is hardly any test of the AdS/CFT duality at
finite $N$.
On the one hand, the $1/N$ corrections are quantum gravity loop
corrections and thus very hard to compute on the gravity side.
On the other hand, there are indications from the field theory
analysis that the integrability breaks down at the non-planar level
\cite{Beisert:2003tq}.
So there was virtually no technique to evaluate the $1/N$ corrections
at strong 't Hooft coupling.

\bigskip
However, there has been important progress in the localization
technique in supersymmetric gauge theories \cite{Pestun:2007rz},
elucidating ealier works \cite{Erickson:2000af,Drukker:2000rr}.
This provides a powerful method to compute the $1/N$ corrections at
strong 't Hooft coupling in dual CFT's. In \cite{Kapustin:2009kz} the
localization technique was applied to ${\cal N}=6$ $U(N)_k\times
U(N)_{-k}$ Chern-Simons-matter theory, also known as the ABJM theory,
dual to type IIA strings on $AdS_4\times CP^3$ or M-theory on
$AdS_4\times S^7/Z_k$ \cite{Aharony:2008ug}.
The partition function of the ABJM theory reduces to the eigenvalue
integrals and this defines the ABJM matrix model.
Subsequently, the free energy of the ABJM matrix model was computed
exactly in the large $N$ limit and order by order in the $1/N$
expansion \cite{Drukker:2010nc,Marino:2011nm,Drukker:2011zy}.
In particular, the genus zero free energy at strong 't Hooft coupling
precisely agrees with the tree level SUGRA action.
This provided a remarkable check of the AdS/CFT duality and showed the
power of the localization technique.\footnote{See related interesting
  developments; generalizations to  quiver gauge theories
  \cite{triSE}, the $F$-maximization conjecture \cite{Herzog:2010hf},
  computations of superconformal index \cite{Imamura:2011su}, and
  tests of three-dimensional dualities \cite{Kapustin:2010xq}.}

\bigskip
In this note we study further the free energy of the ABJM theory at
strong 't Hooft coupling at finite $N$ with the aim of testing AdS/CFT
at the quantum gravity level and, in particular, resum all the $1/N$
corrections.
Our main result is that, apart from the worldsheet instanton
contributions, the all genus free energy of the ABJM matrix model sums
up to 
\be
F^{\rm ABJM}=\log\left[2\pi C_1
\mbox{Ai}\left(\left[{\pi\over\sqrt{2}}\left({N\over\lambda}\right)^2
\lambda_{\rm ren}^{3/2} \right]^{2/3}\right)\right]\ ,
\label{allgenus}
\ee
where $C_1=(2\pi/k)^{-1/3}/\sqrt{2}$ and $\lambda=N/k$ is the 't Hooft
coupling of ${\cal N}=6$ $U(N)_k\times U(N)_{-k}$ Chern-Simons-matter
theory, and
\be
\lambda_{\rm ren}=\lambda-{1\over 24}-{\lambda^2\over 3N^2}\ .
\label{lambdaren}
\ee
This non-planar shift/renormalization of the 't Hooft coupling
(\ref{lambdaren}) partially resums the all genus free energy
\cite{Drukker:2011zy}.
As we will see, once this partial resummation is done, the remaining
higher genus free energies obey a very simple recursion relation
\cite{Huang:2009md} which one can easily solve to find the main result
(\ref{allgenus}).

\bigskip
It is worthwhile to note that the shift (\ref{lambdaren}) is closely
related to the renormalization of the AdS radius (in the $\alpha'=1$
unit) \cite{Bergman:2009zh}:
\be
R_{AdS}^2=2^{5/2}\pi\sqrt{\lambda-{1\over 24}+{\lambda^2\over 24N^2}}\ .
\label{BH}
\ee
In the planar limit $N\to\infty$, the matrix model shift
(\ref{lambdaren}) agrees with that of the dual string theory
(\ref{BH}), as noted in
\cite{Drukker:2010nc,Marino:2011nm}.\footnote{In \cite{Drukker:2010nc}
  the unequal rank $U(N_1)\times U(N_2)$ case \cite{Aharony:2008gk}
  was also studied, and the generalization of (\ref{lambdaren})
  precisely agrees with the SUGRA prediction \cite{Aharony:2009fc} in
  the large $N$ limit.}
However, there appears to be a discrepancy at the non-planar level;
The matrix model suggests a further shift $-3\lambda^2/8N^2$ to the
string/M theory prediction.
This may provide a simplest test of AdS/CFT at the quantum gravity
level, and it is very important to understand whether and how this
discrepancy can be reconciled.
We will return to this point later in the discussions.

\bigskip
The organization of the rest of our paper is as follows: 
In section \ref{ABJMMM} we briefly review the localization technique
and the ABJM matrix model and summarize the result of
\cite{Drukker:2010nc} relevant to our study.
In section \ref{HigherGenus} we give a brief review of the technique
to compute the higher genus free energies by using the so-called
holomorphic anomaly equation and carry out a partial resummation of
the all genus free energy.
In section \ref{AllGenus} we propose the all genus free energy up to
the worldsheet instanton corrections and give a proof of our
proposal.
In section \ref{Discussions} we summarize our results and discuss the
SUGRA one-loop corrections to the free energy.
In the appendices we give the technical details of our computations.

\section{The ABJM matrix model}
\label{ABJMMM}

The localization technique \cite{Pestun:2007rz} allows us to compute
the free energy of the ABJM theory from weak to strong 't Hooft
coupling $\lambda=N/k$ at finite $N$.
The localization of this theory on $S^3$ was carried out by Kapustin,
Willet, and Yaakov in \cite{Kapustin:2009kz};
The partition function reduces to the finite dimensional integral over
the eigenvalues of two $U(N)$ matrices.
This defines the ABJM matrix model.
As demonstrated by Drukker, Mari\~no, and Putrov
\cite{Drukker:2010nc,Drukker:2011zy,Marino:2011nm}, the free energy
can be calculated for arbitrary 't Hooft coupling $\lambda$ and order
by order in the $1/N$ expansion.

\bigskip
The field content of the ABJM theory consists of two 3d ${\cal N}=2$
$U(N)$ vector multiplets,
$(A^A_{\mu},\sigma^A,\lambda^A,\bar{\lambda}^A,D^A)_{A=1,2}$, which
are the dimensional reduction of 4d ${\cal N}=1$ vector multiplets and
two bifundamental chiral multiplets,
$(\phi^I,\bar{\phi}^I,\psi^I,\bar{\psi}^I, F^I,\bar{F}^I)_{I=1,2}$, in
the representation $(N,\bar{N})$, and their duals in
$(\bar{N},N)$.\footnote{In Euclidean space, the barred fields are not
  complex conjugate of the unbarred fields but independent of them.}
This theory enjoys the nilpotent Grassmann-odd symmetry generated by
$\bar{\delta}=\bar{\epsilon} \bar{Q}$ where $\bar{\epsilon}$ being the
two-component complex Killing spinor on $S^3$ with the normalization
$\bar{\epsilon}\epsilon=1$ and $\bar{Q}$ is the ${\cal N}=1$
supercharge.

\bigskip
The partition function is invariant under the deformation of the
action by any $\bar{\delta}$-exact terms, since the ${\cal N}=6$
$U(N)_k\times U(N)_{-k}$ Chern-Simons-matter action and the vacua are
invariant under the $\bar{\delta}$-transformation.
This has an important ramification: By deforming the action by  $t
\bar{\delta} V$ with a positive definite $\bar{\delta} V$, as we send
$t$ to infinity, the path integral localizes on the saddle point of
$\bar{\delta} V$. A particularly convenient choice of
$\bar{\delta}V=\bar{\delta} V_{\rm gauge}+\bar{\delta} V_{\rm matter}$
is \cite{Kapustin:2009kz,Hama:2010av}
\begin{align}
\bar{\delta} V_{\rm gauge}
&=\bar{\delta}\delta\Tr_{U(N)}
\left(\half\bar{\lambda}^A\lambda^A-2D^A\sigma^A\right)=S_{\rm YM}\ ,\\
\bar{\delta} V_{\rm matter}&=\bar{\delta}\delta
\left(\bar{\psi}^I\psi^I
-2i\bar{\phi}^I\left(\sigma^1-\sigma^2\right)\phi^I
-\bar{\phi}^I\phi^I\right)
=S_{\rm m}\ ,
\end{align}
where $S_{\rm YM}$ and $S_{\rm m}$ are the ${\cal N}=2$ super
Yang-Mills action and the bifundamental matter action on the unit
sphere, respectively. Their saddle points are given by
\cite{Kapustin:2009kz}
\be
A^A_{\mu}=\phi^I=0\ ,\qquad
D^A=-\sigma^A=\mbox{const}\ .
\ee
The localization action $S_{\rm YM}+S_{\rm m}$ vanishes on the saddle
points. Thus, in the $t\to\infty$ limit, the only contribution comes
from (1) the classical Chern-Simons-matter action $S_{cl}$ evaluated
on the saddle points and (2) the quadratic fluctuations about the
saddle points in the localization action. The latter gives the
one-loop determinants, and thus schematically the partition function
becomes
\be
Z^{\rm ABJM}=\int d\sigma^1d\sigma^2
{\det\Delta_F(\sigma^A)\over\det\Delta_B(\sigma^A)}
\exp\left(-S_{\rm cl}(\sigma^A)\right)\ ,
\ee
where $\Delta_B$ and $\Delta_F$ denote the Laplacians on $S^3$ for the
bosonic and fermionic fluctuations, respectively.

\bigskip
Diagonalizing the $U(N)$ matrices $\sigma^{i=1,2}$ and integrating
their angular parts yields the Vandermonde determinants which are
cancelled by the factors from the one-loop determinants.
The net result is given by
\cite{Kapustin:2009kz,Marino:2009jd,Drukker:2010nc,Marino:2011nm}
\begin{align}
Z^{\rm ABJM}={1\over \left(N!\right)^2}
\int\prod_{i=1}^N{d\mu_i\over 2\pi}\prod_{a=1}^N{d\nu_a\over 2\pi}
&{\prod_{i<j}\left(2\sinh\left({\mu_i-\mu_j\over 2}\right)\right)^2
\prod_{a<b}\left(2\sinh\left({\nu_a-\nu_b\over 2}\right)\right)^2\over
\prod_{i,a}\left(2\cosh\left({\mu_i-\nu_a\over 2}\right)\right)^2}\nn\\
&\quad\times e^{-{1\over 2g_s}\left(\sum_i\mu_i^2-\sum_a\nu_a^2\right)}\ ,
\label{ABJMmatrixmodel}
\end{align}
where $g_s=2\pi i/k$. $\mu_i$'s and $\nu_a$'s are the eigenvalues of
the two $U(N)$ matrices $\sigma_i$, and the hyperbolic functions are
the one-loop determinant contributions.
Those in the numerator are from the vector multiplets, whereas those
in the denominator are from the bifundamental matter multiplets which
give the coupling between the two $U(N)$ factors.
The eigenvalue integral (\ref{ABJMmatrixmodel}) defines the ABJM
matrix model.

\bigskip
The planar limit of the ABJM matrix model can be solved by the
standard technique
\cite{Drukker:2010nc,Drukker:2011zy,Marino:2011nm}.
In the large $N$ limit, the partition function (\ref{ABJMmatrixmodel})
is dominated by the saddle points:
\begin{align}
\mu_i&={t_1\over N_1}\sum_{j\ne i}^{N_1}\coth{\mu_i-\mu_j\over 2}
+{t_2\over N_2}\sum_{a=1}^{N_2}\tanh{\mu_i-\nu_a\over 2}\ ,
\label{saddle1}\\
\nu_a&={t_2\over N_2}\sum_{b\ne a}^{N_2}\coth{\nu_a-\nu_b\over 2}
+{t_1\over N_1}\sum_{i=1}^{N_1}\tanh{\nu_a-\mu_i\over 2}\ ,
\label{saddle2}
\end{align}
where $t_i=g_sN_i$ and $N_1=-N_2=N$.
In solving these equations, one first considers the problem for
positive $N_1$ and $N_2$ and then analytically continues the result to
the negative $N_2=-N_1=-N$.

\bigskip
The effective potential which leads to the above saddle point
equations consists of attractive harmonic potentials and logarithmic
Coulomb repulsions for the eigenvalues $\mu_i$'s and $\nu_a+i\pi$'s.
Since the Coulomb repulsions have the strength $t_i$'s, when the 't
Hooft couplings $t_i$'s are zero, the harmonic potentials dominate and
the eigenvalues collapses to zero.
As we increase $t_i$'s, the Coulomb repulsions kick in and spread the
eigenvalues along two line intervals;
The $\mu_i$'s condense in ${\cal C}_1$ on the real axis, and the
$\nu_a+i\pi$'s in ${\cal C}_2$, the interval separated from
${\cal C}_1$ by $+i\pi$.
Thus the ABJM matrix model is a two-cut model.

\bigskip
Similarly to the case of the standard matrix models, one defines the
resolvent by
\be
\omega(z)=g_s\left\langle\sum_{i=1}^{N_1}\coth{z-\mu_i\over 2}\right\rangle
+g_s\left\langle\sum_{a=1}^{N_2}\tanh{z-\nu_a\over 2}\right\rangle\ .
\label{resolvent}
\ee 
To make more direct contact with the standard matrix models, it is
useful to introduce the new variable $Z=e^z$.
Then the resolvent is expressed as
\be
\omega(z)dz=-t{dZ\over Z}
+2g_s\left\langle\sum_{i=1}^{N_1}{dZ\over Z-e^{\mu_i}}\right\rangle
+2g_s\left\langle\sum_{a=1}^{N_2}{dZ\over Z+e^{\nu_a}}\right\rangle\ ,
\label{resolventoneform}
\ee
where $t=t_1+t_2$.
As studied in \cite{Halmagyi:2003ze}, the saddle point equations
(\ref{saddle1}) and (\ref{saddle2}) imply that the discontinuity of
the resolvent  $\omega_0(z)\equiv\lim_{N\to\infty}\omega(z)$ at large
$N$ is given by
\begin{align}
z&=\half\left(\omega_0(z+i\epsilon)+\omega_0(z-i\epsilon)\right)
\quad\mbox{on}\quad {\cal C}_1\ ,\\
z&=\half\left(\omega_0(z+i\pi+i\epsilon)+\omega_0(z+i\pi-i\epsilon)\right)
\quad\mbox{on}\quad {\cal C}_2\ .
\end{align}
It then follows that the function
\be
f(Z)=e^t\left(e^{\omega_0}+Z^2e^{-\omega_0}\right)\label{regularfunc}
\ee
is regular everywhere on the complex plane. This has the asymptotic
behavior $f(Z)\stackrel{Z\to\infty}{\longrightarrow} Z^2$ and
$f(Z)\stackrel{Z\to0}{\longrightarrow} 1$.
With these boundary conditions, the function $f(Z)$ is uniquely
determined:
\be
f(Z)=Z^2-\zeta Z+1\ .
\ee
Meanwhile, from (\ref{regularfunc}) one has
\be
\omega_0(Z)=\log\left[{e^{-t}\over 2}
\left(f(Z)-\sqrt{f(Z)^2-4e^{2t}Z^2}\right)\right]\ .
\ee
Parameterizing the branch cuts ${\cal C}_1$ and ${\cal C}_2$ on the
$Z$-plane by $[1/a,a]$ and $[-b,-1/b]$, respectively, the inside of
the square root can be written as the polynomical
$(Z-a)(Z-1/a)(Z+b)(Z+1/b)$, and one can then identify
\be
\zeta=\half\left(a+{1\over a}-b-{1\over b}\right)\ ,\qquad
e^t=\frac{1}{4}\left(a+{1\over a}+b+{1\over b}\right)\ .
\ee

\bigskip
Once the resolvent is found, the 't Hooft couplings $t_i$ can be
computed from
\be
t_i={1\over 4\pi i}\oint_{A_i}\omega_0(z)dz\ ,\label{thooft}
\ee 
and the genus zero free energy from
\be
{\del F_0\over\del s}=\pi i t-\half\oint_{B}\omega_0(z)dz\ ,
\label{free-energy-resolvent}
\ee
where $s=\half(t_1-t_2)$. 
The contour $A_i$ encircles the branch cut ${\cal C}_i$, whereas the
contour $B$ is the cycle dual to the $A_i$ cycles.
The first term in the RHS of (\ref{free-energy-resolvent}) is due to
the pole at $Z=0$ in (\ref{resolventoneform}).
These contour integrals (\ref{thooft}) and
(\ref{free-energy-resolvent}) are hard to carry out.
However, once the derivative is taken w.r.t. $\zeta$, it becomes easy
to perform the integration.

\bigskip
After the analytic continuation
$t_2\to-t_1=-2\pi i{N\over k}= 2\pi i\lambda$, the answer for the ABJM
matrix model turns out to be
\cite{Drukker:2010nc,Drukker:2011zy,Marino:2011nm}
\begin{align}
\lambda(\kappa)&={\kappa\over 8\pi}
{}_3F_2\left(\half,\half,\half;1;{3\over 2};-{\kappa^2\over 16}\right)\ ,\\
\del_{\lambda}F_0(\kappa)&={\kappa\over 4}G^{2,3}_{3,3}
\left(
\begin{array}{ccc}
\!\!\half, &\!\!\half, &\!\!\!\!\half \\
\!\!0, & \!\!0, &\!\!\!\! -\half
\end{array}
\Biggr|\!-{\kappa^2\over 16}
\right)+{\pi^2 i\kappa\over 2}
{}_3F_2\left(\half,\half,\half;1;{3\over 2};-{\kappa^2\over 16}\right)
\end{align}
where $\kappa=-i\zeta$.
In particular, at strong coupling $\lambda \gg 1$, these yield
\begin{align}
\lambda-{1\over 24}&={\log^2\kappa\over 2\pi^2}+{\cal O}(1/\kappa^2)\ ,\\
F_0(\lambda)&={4\sqrt{2}\pi^2\over 3}\left(\lambda-{1\over 24}\right)^{3/2}
+{\cal O}\left(e^{-2\pi\sqrt{\lambda-{1\over 24}}}\right)\ .
\end{align}
Quite remarkably, the first equation shows that, at least in the large
$N$ limit, the free energy of the ABJM theory at strong coupling is
naturally expressed in terms of the shifted 't Hooft coupling
$\lambda-{1\over 24}$, rather than $\lambda$, in agreement with the
SUGRA prediction \cite{Bergman:2009zh}.
Moreover, the first term in the genus zero free energy precisely
agrees with the (minus of) classical SUGRA action.
As $\lambda = N/k$, this in particular reproduces the tantalizing
$N^{3/2}$ scaling of the M2-brane theory.
Lastly, the exponential corrections to the free energy agree with the
expected worldsheet instanton corrections.

\bigskip
These results proclaim the power and relevance of the localization
technique to the test of the AdS/CFT conjecture.
With this remarkable success, it is natural to ask if the AdS/CFT can
be tested beyond large $N$ limit by studying the non-planar
corrections in the ABJM matrix model.

\section{Higher genus free energies}
\label{HigherGenus}

The ABJM matrix model is equivalent to the so-called Lens space matrix
model \cite{Marino:2002fk,Aganagic:2002wv}, which computes the
partition function of the $U(N_1)\times U(N_2)$ branch of the
Chern-Simons theory on the Lens space $L(2,1)=S^3/Z_2$, by the
analytic continuation $N_2\to -N_2$ \cite{Marino:2009jd}.

\bigskip
Meanwhile, as discussed in \cite{Aganagic:2002wv},  the Lens space
matrix model arises as the low energy effective theory of D-branes
wrapping $S^3/Z_2$ in the topological A-model on
$T^{\ast}\!\left(S^3/Z_2\right)$.
At large $N$ the geometric transition takes place and $S^3/Z_2$ is
replaced by the Hirzebruch surface
$\mathbb{F}_0=\mathbb{P}^1\times\mathbb{P}^1$, yielding the
topological A-model on a non-compact Calabi-Yau space, the canonical
line bundle over $\mathbb{F}_0$.
This is the large $N$ (open/closed string) duality between the Lens
space matrix model and the topological A model on local
$\mathbb{F}_0$.
Furthermore, the A-model on local $\mathbb{F}_0$ can be mapped to the
B-model on the mirror manifold given by the surface $uv = H(x,y)$,
where $H(x,y)=0$ is an elliptic curve and coincides with the spectral
curve of the Lens space matrix model.
The closed string amplitudes $F^{(g)}(t,\bar{t})$ of the B-model,
where $t$ is the complex structure moduli, obey the holomorphic
anomaly equation of \cite{Bershadsky:1993cx} and can be explicitly
calculated.
In \cite{Dijkgraaf:2002fc} it was conjectured that the holomorphic
limit $F^{(g)}(t)\equiv\lim_{\bar{t}\to\infty}F^{(g)}(t,\bar{t})$ of
the amplitudes are given by the genus $g$ free energies of the matrix
model whose spectral curve is $H(x,y)=0$.
Direct proof was given in \cite{Eynard:2007hf} built on
\cite{Eynard:2007math} that the solution of the matrix model loop
equation \cite{Ambjorn:1992gw} is that of the holomorphic anomaly
equation in the $\bar{t}\to\infty$ limit when the appropriate boundary
conditions are imposed.
Thus one can use the holomorphic anomaly equation to find the genus
$g$ free energies of the ABJM matrix model which is  equivalent to the
Lens space matrix model by a simple analytic continuation.

\subsection{The holomorphic anomaly equation}

Parameterizing the complex structure moduli by the coordinates $t^I$,
the holomorphic anomaly equation takes the form
\cite{Bershadsky:1993cx}: $(g\ge 2)$
\be
\del_{\bar{I}}F_g=\half C_{\bar{I}\bar{J}\bar{K}}e^{2K}G^{J\bar{J}}G^{K\bar{K}}
\left(D_JD_KF_{g-1}+\sum_{r=1}^{g-1}D_JF_rD_KF_{g-r}\right)\ ,\label{BCOV}
\ee
where $G_{I\bar{J}}=\del_I\del_{\bar{J}}K$ is the K\"ahler metric on
the moduli space, $C_{IJK}$ is the Yukawa coupling defined by
$C_{IJK}=\del_I\del_J\del_KF_0$ with $F_0$ being the genus $0$ free
energy, and the covariant derivative $D_I$ acts on $F_g$ and its
derivatives as $\del_I-\Gamma^{\bullet}_{I\bullet}+(2-2g)\del_IK$.
In the local Calabi-Yau case, the K\"ahler potential is given by
$K={i\over 2}\left(t^K\del_{\bar{K}}\bar{F}_0-t^{\bar{K}}\del_KF_0\right)$,
and thus the moduli space metric is $G_{I\bar{J}}=\mbox{Im}\tau_{IJ}$
with $\tau_{IJ}\equiv \del_I\del_JF_0$ \cite{Huang:2009md}.

\bigskip
When applied to the ABJM matrix model \cite{Drukker:2010nc}, the
moduli space becomes one-dimensional and the modulus $t$ is identified
with the 't Hooft coupling $\lambda$.
Then the Yukawa coupling is
$C_{\lambda\lambda\lambda}=4\del_{\lambda}^3F_0(\lambda)
=-32\pi^3i\xi$ where 
\be
\xi={2\over\vartheta_2(\tau)^2\vartheta_4(\tau)^4}\ ,\qquad
\tau={i\over 4\pi^3}\del_{\lambda}^2F_0(\lambda)+1
=i{K'\left({i\kappa\over 4}\right)\over K\left({i\kappa\over 4}\right)}
\label{tau}
\ee
with the normalizations properly adjusted.
The genus $g$ free energies are assumed to be of the form
\begin{align}
F_g(\tau)&=\xi^{2g-2}f_g(\tau)\ ,\\
f_g(\tau)&=\sum_{k=1}^{3g-3}{A_g^{(k)}(\tau)\over k}
\left({E_2(\tau)\over 12}\right)^k+c_g^0(\tau)\ ,
\end{align}
where $A_g^{(k)}(\tau)$'s are the modular forms of weight $6g-6-2k$,
and $c_g^0(\tau)$ is the holomorphic ambiguity and a modular form of
weight $6g-6$.
The Eisenstein series of weight $2$, $E_2(\tau)$, is a quasi-modular
form and can be promoted to the non-holomorphic modular form as
\be
E_2(\tau)\longrightarrow
\hat{E}_2(\tau,\bar{\tau})=E_2(\tau)-{3\over\mbox{Im}\tau}\ .
\ee
By promoting $F_g(\tau)$ to the non-holomorphic modular form
$F_g(\tau,\bar{\tau})$ in this way, the non-holomorphicity comes in
only through $E_2$.
It then follows from the holomorphic anomaly (\ref{BCOV}) that
$(g\ge 2)$
\be
{df_g\over dE_2}=-\frac{1}{3}\left\{
d^2_{\xi}f_{g-1}+\frac{1}{3}\frac{\partial_{\tau}\xi}{\xi}d_{\xi}f_{g-1}
+\sum_{r=1}^{g-1}d_{\xi}f_rd_{\xi}f_{g-r}
\right\}\ ,\label{HAE}
\ee
where the covariant derivative $d_{\xi}$ transforms a form of weight
$k$ to a form of weight $k+2$:
$d_{\xi}=\del_{\tau}+{k\over 3}{\del_{\tau}\xi\over\xi}$.
In Appendix A we collect various formulae used in the subsequent
computations.

\bigskip
The genus one free energy can be independently calculated by using
Akemann's formula \cite{Akemann:1996zr}.
One can find that
\be
F_1(\tau)=-\log\eta(\tau)\ ,
\ee
where $\eta(\tau)$ is the Dedekind eta function.
This is the initial data needed to solve the holomorphic anomaly
equation.
In particular, at strong coupling $\lambda\sim \log^2\kappa\gg 1$,
this is expanded as
\be
F_1(\tau)={1\over 6}\log\kappa-\half\log\left(\log\kappa\right)
+{\cal O}(1/\kappa)\ .
\label{torusFE}
\ee

\subsection{The weight zero free energy}

We first focus on $A_g:=A_g^{(3g-3)}$'s of modular weight zero.
By using the formulae in Appendix A, it is easy to show that they obey
a simple recursion relation \cite{Huang:2009md}
\be
{A_g\over 12}=-{1\over 3}\left((3g-3)A_{g-1}
+\sum_{r=2}^{g-2}A_rA_{g-r}\right)\ .\label{leadingrecursion}
\ee
They are the highest order terms in $F_g$'s in the $x=1/\log\kappa$
expansion.
Indeed, using the modular transformation properties in Appendix A, one
finds that
\be
F_g={(-1)^{g-1}\over 4^{2g-2}}{A_g\over 3g-3}x^{3g-3}
+{\cal O}\left(x^{3g-4}\right)
:=F^{[0]}_gx^{3g-3}+{\cal O}\left(x^{3g-4}\right)\ ,\label{highestFg}
\ee
where $x=1/\log\kappa$.
As it will turn out, the leading order recursion
(\ref{leadingrecursion}) carries almost all the information needed to
sum up the all genus free energy, when the worldsheet instanton
corrections are ignored.

\bigskip
We can in fact resum the highest order all genus free energy
\be
F(t)=\sum_{g=2}^{\infty}F^{[0]}_g t^{2g-2}\ ,
\ee 
where $t=-ig_sx^{3/2}$.
For the convenience, we instead consider the generating function
\be
f(u):=\sum_{g=2}^{\infty}\tilde{A}_g u^{g}\ ,
\ee
where $\tilde{A}_g=(g-1)(4/3)^{g-1}F^{[0]}_g$.
Then it is easy to find that
\be
f(u)-\tilde{A}_2u^2=u^2f(u)'+f(u)^2 \label{gen1}
\ee
with $\tilde{A}_2=5/36$ which can be found from the initial genus one
data.
This can be solved to
\be
f(u)=-{d\over du^{-1}}\log H(u)\ ,\label{generating1}
\ee
where 
\be
H(u)=e^{-{1\over 2u}}
\sqrt{1/2u}\left(C_1K_{1/3}(1/2u)+C_2I_{1/3}(1/2u)\right)\ ,\label{H}
\ee
with $C_1$ and $C_2$ being integration constants. 
The highest order all genus free energy can then be computed as
\be
F(t)=-\int^{4/3t^2}du^{-1}f(u)=\log H(3t^2/4)\ .
\ee
Note that we absorbed the integration constant into $C_1$ and $C_2$. 

\bigskip
To fix the constants $C_1$ and $C_2$, we require the absence of the
non-perturbative corrections of the type $e^{-2/3t^2}$ for small $t$.
Since $t\sim g_s$, this is of the order $e^{-1/g_s^2}$ and would be
the gravitational instanton or NS5-brane effect, as opposed to the
D-brane effect, in the dual AdS string theory.
We assume that the type IIA string theory on $AdS_4\times CP^3$ does
not receive such corrections.
With this assumption, there remain two choices;
(1) $C_2=0$ or (2) $C_1=-(-1)^{5/6}C_2/\pi$.
However, the latter choice is subject to a further non-perturbative
ambiguity due to the Stokes phenomenon of the modified Bessel function
$I_{1/3}(z)$.
So we claim that the correct choice is $C_2=0$.
The remaining constant $C_1$ only shifts the constant part of the
genus one free energy and is not physically important.

\bigskip
Using the identity $\mbox{Ai}(z)
={1\over\pi}\sqrt{z\over 3}K_{1/3}\left({2\over 3}z^{3/2}\right)$, we
find
\be
F\left(t\right)
=\log\left[2\pi C_1 e^{-{2\over 3t^2}}t^{-{1\over 3}}\mbox{Ai}(t^{-4/3})\right]\ ,
\ee
where $t=-ig_sx^{3/2}$.
Note that the factors in front of the Airy function are precisely the
minus of the highest order terms (in $x$) of the genus zero and one
contributions.
Hence the total weight zero free energy becomes simply
\be
\widehat{F}^{[0]}\left(g_s,x\right)
=\log\left[2\pi C_1
\mbox{Ai}\left(\left(-g_s^2x^3\right)^{-2/3}\right)\right]\ .
\label{weightzeroFE}
\ee
As we will see momentarily, we have actually resummed a significant
part of the all genus free energy up to the worldsheet instanton
corrections.

\section{The all genus free energy}
\label{AllGenus}

It was observed in \cite{Drukker:2011zy} that the all genus free
energy, apart from the worldsheet instanton corrections, depends on
the 't Hooft coupling only through the shifted/renormalized form
\be
\lambda_{\rm ren}=\lambda -{1\over 24} -{\lambda^2\over 3N^2}\ .
\ee
This translates to the renormalization of $g_sx$ by
\be
g_sx\to g_sy\equiv{g_sx\over\sqrt{1+(g_sx)^2/6}}\ ,
\ee
where $g_s=2\pi i \lambda/N$ and
$x^{-1}=\sqrt{2}\pi\sqrt{\lambda-1/24}$.

\bigskip
We now claim that, apart from the worldsheet instanton corrections,
the all genus free energy can be obtained by replacing $x$ with the
renormalized variable $y$ in the weight zero free energy
(\ref{weightzeroFE}).
That is, the all genus free energy of the ABJM theory is given by
\be
F^{\rm ABJM}\left(\lambda,N\right)
=\log\left[2\pi C_1
\mbox{Ai}\left(\left(-g_s^2y^3\right)^{-2/3}\right)\right]
+{\cal O}\left(e^{-2\pi\sqrt{\lambda-{1\over 24}}}\right)\ ,\label{allgenusFE}
\ee
where $C_1=(-g_s^2)^{-1/6}/\sqrt{2}$ to match the normalization in
\cite{Drukker:2011zy}.
This is the main result (\ref{allgenus}).

\bigskip
To show it, we first note that, when the worldsheet instanton
corrections, {\it i.e.}, the ${\cal O}(1/\kappa)$ terms, are
neglected, the holomorphic anomaly equation (\ref{HAE}) becomes
\begin{eqnarray}
F^{\prime}_g(x)
=\frac{1}{4}x^4F^{\prime\prime}_{g-1}(x)
+\frac{12x-1}{12}x^2F^{\prime}_{g-1}(x)
+\frac{x^4}{4}\sum_{r=2}^{g-2}F^{\prime}_r(x)F^{\prime}_{g-r}(x)\ .
\label{recursion2}
\end{eqnarray}
To derive this, we performed the modular transformation
$ST^{-1}:\tau\mapsto\tau'=-1/(\tau-1)$, and kept only the terms in
powers of $\log\kappa=\pi i\tau'/2
+{\cal O}\left(e^{2\pi i\tau'}\right)$.
We then used the formulae listed in Appendix A.

\bigskip
To proceed, we define the generating function
\be
{\cal F}(g_s, x):=\sum_{g=2}^{\infty}g_s^{2g-2}F_g'(x)\label{generating}
\ee
which obeys
\begin{eqnarray}
{\cal F}(g_s,x)-g_s^2F_2^{\prime}(x)
={1\over 4}g_s^2x^4\partial_x{\cal F}(g_s,x)
+\frac{12x-1}{12}g_s^2x^2{\cal F}(g_s,x)
+{1\over 4}g_s^2x^4{\cal F}(g_s,x)^2\ ,
\label{diff_rec}
\end{eqnarray}
where 
\begin{eqnarray}
F_2(x)=\frac{x}{144}-\frac{x^2}{24}+\frac{5x^3}{48}\ .
\end{eqnarray}
Then the claim (\ref{allgenusFE}) is equivalent to
\begin{eqnarray}
{\cal F}(g_s,x)=\frac{\del y}{\del x}
\left\{{4\over g_s^2y^4}\widehat{f}\left(\frac{3}{4}g_s^2y^3\right)
-\partial_y\left[\frac{2}{3g_s^2x^3}
+\left(\frac{1}{6x}+\frac{1}{2}\log x\right)\right]\right\}\ ,
\label{conjecture0}
\end{eqnarray}
where
\be
\widehat{f}(z):=f(z)+\left[-\half +{z\over 6}\right]\ .
\ee
is the generating function (\ref{generating1}) plus the genus zero and
one contributions, while the second term in (\ref{conjecture0}) is the
subtraction of the genus zero and one contributions.
Plugging (\ref{conjecture0}) into (\ref{diff_rec}) yields
\be
f\left(\frac{3}{4}g_s^2y^3\right)
-\frac{5}{36}\left(\frac{3}{4}g_s^2y^3\right)^2
=\left(\frac{3}{4}g_s^2y^3\right)^2f^{\prime}\left(\frac{3}{4}g_s^2y^3\right)
+f\left(\frac{3}{4}g_s^2y^3\right)^2\ .
\ee
This is equivalent to (\ref{gen1}).
Hence we have shown that the all genus free energy (\ref{allgenusFE})
indeed satisfies the holomorphic anomaly equation (\ref{HAE}) up to
the worldsheet instanton corrections.

\section{Discussions and conclusions}
\label{Discussions}

We succeeded to sum up the all genus free energy of the ABJM theory,
apart from the worldsheet instanton corrections.
The resummation was done in two steps;
(1) First, we summed up the modular weight zero part of the all genus
free energy as given in (\ref{weightzeroFE}).
(2) Then the remaining parts were resummed to (\ref{allgenusFE}) by
renormalizing the 't Hooft coupling as was done in
\cite{Drukker:2011zy}.
The renormalization (\ref{lambdaren}) agrees with the SUGRA prediction
\cite{Bergman:2009zh} in the large $N$ limit.
However, there seems to be a discrepancy at the non-planar level;
The matrix model suggests a further shift $-3\lambda^2/8N^2$ to the
string/M-theory prediction (\ref{BH}).

\bigskip
The quantum gravity one-loop shift $+\lambda^2/24N^2$ of the AdS
radius in (\ref{BH}) accounts for, at the least, a part of the
one-loop corrections.
This comes from the shift of the M2/D2-brane charge
\cite{Bergman:2009zh}
\be
N\to N-{1\over 24}\left(k-{1\over k}\right)=k\left(\lambda -{1\over 24}
+{\lambda^2\over 24N^2}\right)\ ,\label{chargeshift}
\ee
due to the higher curvature correction $C_3\wedge I_8$, where $I_8$ is
a curvature $8$-form anomaly polynomial \cite{Duff:1995wd}.
One might wonder if the one-loop discrepancy $-3\lambda^2/8N^2$ comes
from other SUGRA one-loop corrections such as the $R^4$ term
\cite{Green:1997di}.\footnote{The $\lambda^2/N^2$ shift appears of the
  order $\sqrt{\lambda}$ in the genus one free energy.}
However, we now argue that they are absent:
In fact, the one-loop $R^4$ term is nonzero and of the order
$\sqrt{\lambda}$ which agrees with the leading part of the genus one
free energy in (\ref{torusFE}).
However, the numerical factor does not agree with the
$-3\lambda^2/8N^2$ discrepancy, and the $R^4$ term is not the only
one-loop correction.
The one-loop $R^4$ term is expected to be completed by the terms
involving the 4-form $F_4$.
Indeed, the 11d ${\cal N}=1$ SUGRA on-shell superfield
\cite{Cremmer:1980ru} suggests schematically the completion of the
form $\left(R+F_4^2\right)^4$, but this vanishes on
$AdS_4\times S^7/Z_k$ \cite{Kallosh:1998qs,Tseytlin:2000sf}.
Furthermore, the one-loop vacuum energy vanishes in 4d ${\cal N}\ge 5$
SUGRA on the global $AdS_4$ whose boundary is $\mathbb{R}\times S^2$
\cite{Allen:1983an}.
Although we work in the Euclidean $AdS_4$ with the boundary $S^3$, the
localizations of the ABJM theory on $S^3$ and $\mathbb{R}\times S^2$
turn out to be the same \cite{Hama:2011ea}.
So we expect the SUGRA result on the global $AdS_4$ applies to our
case, and it seems likely that the one-loop corrections, except for
the charge shift (\ref{chargeshift}), are absent.

\bigskip
This leaves us the $-3\lambda^2/8N^2$ discrepancy and suggests us to
look for the resolution on the matrix model side. 
Recall that ABJM proposed two ${\cal N}=6$ superconformal
Chern-Simons-Matter theories \cite{Aharony:2008ug};
One has the gauge group $U(N)\times U(N)$ and the other has
$SU(N)\times SU(N)$.
In the large $N$ limit, their difference may not matter.
But at finite $N$ they will differ from each other in the $1/N$
corrections.
In fact, in the $AdS_5\times S^5$ case, the correct gauge group is
$SU(N)$ rather than $U(N)$ \cite{Witten:1998qj}.
Although the status of this subtlety is unclear in the $AdS_4$ case,
it may be worthwhile to investigate $SU(N)\times SU(N)$ ABJM matrix
model to see if the $-3\lambda^2/8N^2$ discrepancy can be resolved. We
hope to address this issue further in the near future.

\bigskip
Finally, it is worth emphasizing the remarkable simplicity of our
result;
The partition function of the ABJM theory, when the worldsheet
instantons are neglected, is simply the Airy function.
This might suggest a possible connection of the ABJM theory to the
Kontsevich matrix model \cite{Kontsevich:1992ti} upon the inclusion of
the worldsheet instanton corrections.
Meanwhile, in a somewhat different context of M-theory flux
compactification, it was argued that the norm square of the
``wave-function of the universe'' or the 5d black hole entropy was
given by the Airy function \cite{Ooguri:2005vr}.\footnote{The authors
  are grateful to Hirosi Ooguri for pointing this out to us. See also
  \cite{Marinotalk}.}
It would be interesting to study the relations to these works in the
future.

\section*{Acknowledgment}

We would like to thank Oren Bergman, Nadav Drukker, Kazuo Hosomichi,
Hiroaki Kanno, Masahide Manabe, Kazuhiro Sakai, and Masaki Shigemori
for useful discussions and conversations.
This work was partially supported by the Grant-in-Aid for Nagoya
University Global COE Program (G07), by that for Young Scientists (B)
[\# 21740179] (H.F.) and by [\# 21740176] (S.M.) from the Ministry of
Education, Culture, Sports, Science and Technology of Japan.

\appendix
\renewcommand{\theequation}{\Alph{section}.\arabic{equation}}

\section{Formulae for holomorphic anomaly equation}
In this appendix, we gather various useful formulae related to the
holomorphic anomaly equation:
\begin{align}
&b\equiv\vartheta_2(\tau)^4\ ,\quad
c\equiv\vartheta_3(\tau)^4\ ,\quad
d\equiv\vartheta_4(\tau)^4\ ,
\nn\\
&\frac{\partial_{\tau}\xi}{\xi}=\frac{b-E_2}{4}\ ,\quad
d_{\xi}f_1=-\frac{E_2}{24}\ ,\quad
d_{\xi}b=\frac{1}{3}b(b+d)\ ,\quad
d_{\xi}d=-\frac{1}{6}d(b+d)\ ,
\nn\\
&d_{\xi}E_2=\frac{2bE_2-E_2^2-E_4}{12}\ ,\quad
d_{\xi}E_4=\frac{bE_4-E_6}{3}\ ,\quad
d_{\xi}E_6=\frac{bE_6-E_4^2}{2}\ .
\end{align}
The following formulae are useful to derive the weight zero recursion
(\ref{leadingrecursion})
\begin{align}
f_g&=\frac{A_g^{(3g-3)}}{3g-3}
\left(\frac{E_2}{12}\right)^{3g-3}+\cdots\ ,\nn\\
d_\xi f_g&=-A_g^{(3g-3)}
\left(\frac{E_2}{12}\right)^{3g-2}+\cdots\ ,\nn\\
d_\xi^2f_g&=(3g-2)A_g^{(3g-3)}
\left(\frac{E_2}{12}\right)^{3g-1}+\cdots\ .
\end{align}
To study the perturbative terms as in (\ref{recursion2}), we shall use
the modular transformations $ST^{-1}:\tau\mapsto\tau'=-1/(\tau-1)$,
since (\ref{tau}) implies that $\tau$ is given by
$\tau=1+\pi i/(2\log\kappa)$, neglecting the
${\cal O}(1/\kappa)$ instanton terms, 
\begin{align}
b(\tau)&=-(-i\tau')^2d(\tau')
=(\tau')^2[1+{\cal O}(1/\kappa)]\ ,\nn\\
d(\tau)&=(-i\tau')^2c(\tau')
=(\tau')^2[-1+{\cal O}(1/\kappa)]\ ,\nn\\
E_2(\tau)&=(\tau')^2[E_{2}(\tau')-6i/(\pi\tau')]
=(\tau')^2[(1-3x)+{\cal O}(1/\kappa)]\ ,\nn\\
\xi^2(\tau)&=4/\bigl(b(\tau)d(\tau)^2\bigr)
=(\tau')^{-6}[4+{\cal O}(1/\kappa)]\ ,
\end{align}
with $x=1/\log\kappa$.
Note that since the holomorphic anomaly equation is a modular
covariant equation, the modular scale factors $\tau'$ are destined to
be cancelled out in the end.
We shall drop them in our following analysis.
Then, we obtain the following formulae in the approximation neglecting
the ${\cal O}(1/\kappa)$ terms:
\begin{eqnarray}
d_{\xi}E_2=-\frac{3}{4}x^2\ ,\quad
d^2_{\xi}E_2=-\frac{3}{8}x^3\ ,\quad
\frac{1}{3}\frac{\partial_{\tau}\xi}{\xi}=\frac{x}{4}\ ,\quad
\frac{dE_2}{dx}=-3\ ,\quad
d_{\xi}b=d_{\xi}d=0\ .
\end{eqnarray}
From these we find
\begin{align}
\frac{df_g}{dE_2}&=-\frac{1}{3}f^{\prime}_g(x)\ ,\nn\\
d_{\xi}f_g&=(d_{\xi}E_2)\frac{df_g}{dE_2}=\frac{x^2}{4}f_g^{\prime}(x)\ ,\nn\\
d^2_{\xi}f_g&
=(d_{\xi}^2E_2)\frac{df_g}{dE_2}+(d_{\xi}E_2)^2\frac{d^2f_g}{dE_2^2}
=\frac{x^3}{8}f^{\prime}_{g}(x)+\frac{x^4}{16}f^{\prime\prime}_{g}(x)\ . 
\end{align}

\section{A derivation of weight zero free energy}
In this appendix we solve the differential equation (\ref{gen1}):
Introducing the new variable $v=1/(2u)$ and the new function
\begin{align}
f=-\frac{1}{2}\frac{d}{dv}\log H\ ,
\end{align}
this becomes
\begin{align}
\biggl[\frac{d^2}{dv^2}+2\frac{d}{dv}+\frac{\tilde A_2}{v^2}\biggr]H=0\ .
\end{align}
In terms of $G(v)=e^v v^{-\frac{1}{2}}H(v)$, one obtains $(\alpha=1/3)$
\begin{align}
\biggl[v^2\frac{d}{dv}+v\frac{d}{dv}-\left(v^2+\alpha^2\right)\biggr]G=0\ ,
\end{align}
which is nothing but the modified Bessel's differential equation.
The solution is given by the modified Bessel's functions:
$G(v)=C_1K_{1/3}(v)+C_2I_{1/3}(v)$.

\end{document}